\documentclass[12pt]{iopart}

\usepackage{amsthm, amssymb}

\usepackage{graphicx} 

\usepackage{bm} 

\begin{document}

\title{Anharmonic oscillations of a conical buoy}
\author{}

\author{J Brochado Oliveira, J Monteiro Moreira and J M Machado da Silva}
\address{Departamento de Fisica e Astronomia and IFIMUP - Instituto de Nanomateriais (IN), Universidade do Porto, Portugal}
\ead{jboliv@fc.up.pt}

\date{\today}

\begin{abstract}
A study of the floating of a circular cone shaped buoy in an ideal fluid has revealed some new interesting results. Using reduced variables it is shown, that at a crossover value $(\frac{3}{4})$ of the ratio of the specific masses of the fluid and of the buoy, the anharmonicity of the oscillation is the highest and that, unexpectedly, above this crossover value the normalized period is constant.

\end{abstract}

%\noindent{\it Keywords \/}
%\maketitle

\section{Introduction}

Anharmonic oscillators are useful examples of nonlinear phenomena. Many vibrating systems found in the real world are nonlinear whether they be macroscopic mechanical oscillators \cite{filliponi, arnold, whineray, pecori} or microscopic atomic oscillators\cite{ashcroft}.
The pendulum at high angles is a classic example of an anharmonic oscillator\cite{lewowski, lima}. A particular feature is that the restoring force is equivalent to a spring that softens at large amplitudes i.e.\ $F=-k_1\theta+k_2\theta^3.$\cite{pecori}
In this work we do a detailed analysis of an oscillator materialized by the bobbing cone; we assume that the movement is a vertical translation and that the top of the solid is always emerged and the base is always submerged in an ideal fluid for which the restoring force is due to Archimedes' principle. The restoring force being of the type $F=-a y+b y^2-cy^3~(a, b, c ~\mbox{positive parameters})$ means that the oscillator does not move symmetrically about the origin notwithstanding being periodic.

\section{Restoring force}
\begin{figure}[h!]
\centering
\includegraphics[width=3.0 in]{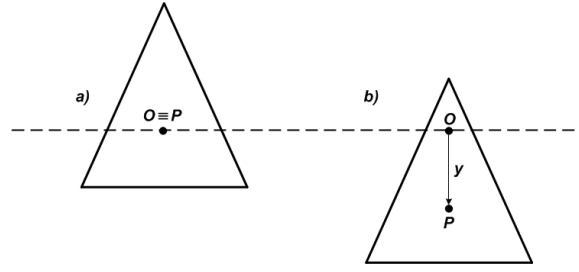}
\caption{The floating cone in two positions: a) equilibrium;    b) at an instant t, with coordinate y. The referential axis is vertical downward.}
\label{fig:figUm}
\end{figure}

Figure\,(\ref{fig:figUm}) schematically represents a floating cone of radius $R$, height $h$ and specific mass $\rho$, partially immersed in an ideal fluid of specific mass $\rho_{\mbox{\scriptsize f}}\,$ ($\rho<\rho_{\mbox{\scriptsize f}}\,$). The position of the solid is chosen to be the coordinate of the point $P$ which coincides with the origin of the reference axis when the buoy is at equilibrium. This origin (point $O$) is the intersection of the axis of the cone at equilibrium with the plane of the free liquid.

The resultant of the forces acting on the solid is $F=\rho V_{\mbox{\scriptsize c}}\,g - \rho_{\mbox{\scriptsize f}}\,(V_0+\Delta V)g$ where $V_0$ is the volume of the immersed cone in static equilibrium ($P\equiv O$ i.e.\ $y=0$), $V_{\mbox{\scriptsize c}}$ is the volume of the cone, $\Delta V$ is the difference between the immersed volume $V$ and $V_0$ and $g$ is the gravity acceleration. 
Therefore:

\begin{eqnarray*}
&\rho V_{\mbox{\scriptsize c}} g= \rho_{\mbox{\scriptsize f}} V_0 g\\
&\Delta V= V_{\mbox{\scriptsize c}}\left[\left(\frac{y}{h}\right)^3-3(1-\alpha)^{1/3}\left(\frac{y}{h}\right)^2+3(1-\alpha)^{2/3}\left(\frac{y}{h}\right)\right]
\end{eqnarray*}

where $\alpha={\rho}/{\rho_{\mbox{\scriptsize f}}}<1.$ %\frac

Then,
\begin{equation*}
F=-\rho_{\mbox{\scriptsize f}}\,g\Delta V=-F_0\left[\left(\frac{y}{h}\right)^3-3(1-\alpha)^{1/3}\left(\frac{y}{h}\right)^2+3(1-\alpha)^{2/3}\left(\frac{y}{h}\right)\right]
\end{equation*}
where $F_0=\rho_{\mbox{\scriptsize f}}\,g V_{\mbox{\scriptsize c}}$.

Introducing the reduced variables $\overline{y}={y}/{h}$ and $\overline{F}={F}/{F_0}$, the reduced restoring force is:
\begin{equation*}
\overline{F}=-\left[\overline{y}^3-3(1-\alpha)^{1/3}\,\overline{y}^2+3(1-\alpha)^{2/3}\,\overline{y}\right]=-P_3(\overline{y},\alpha)
\end{equation*}
where $P_3(\overline{y},\alpha)$ is a third degree polynomial with coefficients that depend on the parameter $\alpha$. 

The reduced coordinates of the vertex and of the center of the base are, at equilibrium, respectively: $\overline{y}_{\mbox{\scriptsize V}}=- \sqrt[3]{1-\alpha}$ and $\overline{y}_{\mbox{\scriptsize C}}=1- \sqrt[3]{1-\alpha}$ . Therefore, due to the restrictions of the movement $-\overline{y}_{\mbox{\scriptsize C}}<\overline{y}<-\overline{y}_{\mbox{\scriptsize V}}$. However, another condition should be imposed on the value of the initial position $y_0$ (or its reduced form $\overline{y}_0$) as a consequence of the potential energy barrier, i.e.\ $E_{\mbox{\scriptsize P}}(y_0)\leq E_{\mbox{\scriptsize P}}(-y_{\mbox{\scriptsize V}})$ and $E_{\mbox{\scriptsize P}}(y_0)\leq E_{\mbox{\scriptsize P}}(-y_{\mbox{\scriptsize C}})$.

\section{Potential energy}
%is given by $$E_{\mbox{\scriptsize P}}=\int_0^y -Fdy$$ where $F$ is the restoring force and 
The potential energy associated with the restoring force $F$, assuming that $E_{\mbox{\scriptsize P}}=0$ at the equilibrium position ($y=0$) is
\begin{eqnarray*}
E_{\mbox{\scriptsize P}}&=&\int_0^y\rho_{\mbox{\scriptsize f}}\, g V_{\mbox{\scriptsize c}}\left[\left(\frac{y}{h}\right)^3-3(1-\alpha)^{1/3}\left(\frac{y}{h}\right)^2+3(1-\alpha)^{2/3}\left(\frac{y}{h}\right)\right]dy\\
&=&\rho_{\mbox{\scriptsize f}}\, g V_{\mbox{\scriptsize c}}\int_0^{\overline{y}} \left[\overline{y}^3-3(1-\alpha)^{1/3}\,\overline{y}^2+3(1-\alpha)^{2/3}\,\overline{y}\right]d\overline{y}.
\end{eqnarray*}
In reduced form, $\overline{E}_{\mbox{\scriptsize P}}={E_{\mbox{\scriptsize P}}}/{E_0}=P_4(\overline{y},\alpha)$, with
\begin{eqnarray*}
&E_0=\rho_{\mbox{\scriptsize f}}\, g h V_{\mbox{\scriptsize c}}=F_0 h\quad \mbox{and}\\
&P_4(\overline{y},\alpha)=\int_0^{\overline{y}} P_3(\overline{y})\,d\overline{y}=\frac{1}{4}\overline{y}^4-(1-\alpha)^{1/3}\,\overline{y}^3+\frac{3}{2}(1-\alpha)^{2/3}\,\overline{y}^2.
\end{eqnarray*}
The choice for the initial values ($y=y_0$ and zero velocity) is subjected to the relations 

\begin{equation}\label{conditons}
\left\{
\begin{array}{l}
-\overline{y}_{\mbox{\scriptsize C}}\leq \overline{y}_0 \leq -\overline{y}_{\mbox{\scriptsize V}} \Longleftrightarrow -1 +\sqrt[3]{1-\alpha}\leq \overline{y}_0\leq \sqrt[3]{1-\alpha}\\\vspace{-5mm}
\\
P_4(\overline{y}_0)= \min\,\left\{\overline{P}_4(-\overline{y}_{\mbox{\scriptsize C}}), \overline{P}_4(-\overline{y}_{\mbox{\scriptsize V}})\right\}
\end{array}\right.
\end{equation}

These equations express the condition that the oscillation has the maximum energy compatible with the fact that the cone is neither completely immersed nor completely emerged.

\begin{figure}[h!]
\centering
\includegraphics[width=4.0 in]{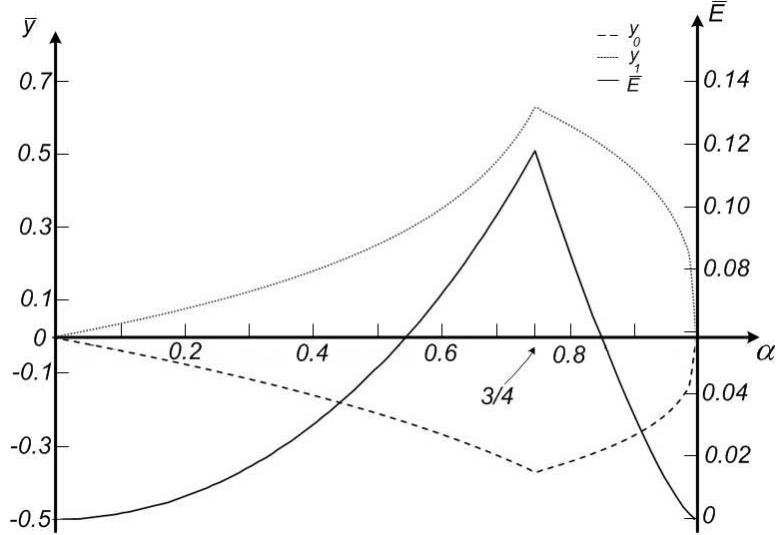}
\caption{The reduced total energy, $\overline{E}=E/E_0$, and the reduced interval, $\overline{y}=y/h$, of the oscillations as a function of $\alpha$. The maximum of the energy $\overline{E}= 3(1/4)^{7/3}$ occurs at the crossover $\alpha=3/4$.}
\label{fig:FigureDoisSemCor}
\end{figure}

It is easy to verify that 
\begin{eqnarray*}
P_4(-\overline{y}_{\mbox{\scriptsize C}})=P_4(-1+\sqrt[3]{1-\alpha})=\frac{3}{4}\,(1-\alpha)^{4/3}+\alpha -\frac{3}{4} \quad \mbox{and}\\
P_4(-\overline{y}_{\mbox{\scriptsize V}})=P_4(\sqrt[3]{1-\alpha})=\frac{3}{4}\,(1-\alpha)^{4/3}.
\end{eqnarray*}
Then
\begin{equation*}
P_4(\overline{y}_0)=\min\left\{\frac{3}{4}\,(1-\alpha)^{4/3}, \frac{3}{4}\,(1-\alpha)^{4/3}+\alpha -\frac{3}{4}\right\};
\end{equation*}
for $\alpha=3/4$ $\Longrightarrow$ $P_4(-\overline{y}_{\mbox{\scriptsize C}})=P_4(-\overline{y}_{\mbox{\scriptsize V}})$. So, if the initial velocity is zero then the initial amplitude $\overline{y}_0$ should be chosen along with
\begin{equation}\label{eqyzero}
\hspace{-1.5cm}\left\{
\begin{array}{l}
\alpha \leq \frac{3}{4} \Longrightarrow \overline{y}_0=-\overline{y}_{\mbox{\scriptsize C}}= -1+\sqrt[3]{1-\alpha}~ \mbox{and}~ \overline{E}_{Total}=\frac{3}{4}\,(1-\alpha)^{4/3}+\alpha -\frac{3}{4}\\\vspace{-5mm}
\\
\alpha > \frac{3}{4} \Longrightarrow \overline{y}_0= -\overline{y}_{\mbox{\scriptsize V}}= \sqrt[3]{1-\alpha}~ \mbox{and}~ \overline{E}_{Total}= \frac{3}{4}\,(1-\alpha)^{4/3}.
\end{array}\right.
\end{equation}
The other limit $\overline{y}_1$ of the interval of $\overline{y}$ can be obtained from the roots of the polynomial $P_4(\overline{y}_1)- {E}_{Total}=0$. There are four solutions (two real and two complex conjugate); the real are the relevant solutions: $\overline{y}_0$ [Eqs.\,(\ref{eqyzero})] and $\overline{y}_1$ [Eqs.\,(\ref{eqyum})]

\begin{equation}\label{eqyum}\small
\hspace{-2.5cm}\left\{
\begin{array}{ll}
\alpha \leq \frac{3}{4} \Longrightarrow \overline{y}_1&=\displaystyle\frac{\left[-44+54\alpha+6\sqrt{-27 + 81\,(1-\alpha)^2 +30\,\alpha}\right]^{1/3}}{3}\,\raisebox{-0.1mm}{\Large -}\\\vspace{-6mm}\vspace{2.5mm}
\\
&\displaystyle -\frac{2}{3\left[-44+54\alpha+6\sqrt{-27 + 81\,(1-\alpha)^2 +30\,\alpha}\right]^{1/3}} + (1-\alpha)^{1/3} + {\displaystyle\frac{1}{3}}\\\vspace{-5mm}
\\
\alpha > \frac{3}{4} \Longrightarrow \overline{y}_1&= -(\sqrt[3]{4}-1)\,\sqrt[3]{1-\alpha}.
\end{array}\right.
\end{equation}
For each $\alpha$ value there is a definite maximum of total energy and a corresponding interval of amplitude $[a,b]\equiv [-\overline{y}_{\mbox{\scriptsize C}}, \overline{y}_1]$ for $\alpha \leq 3/4$ and $[\overline{y}_1, -\overline{y}_{\mbox{\scriptsize V}}]$ for $\alpha > 3/4$ given by Eqs.\,(\ref{eqyzero}) and (\ref{eqyum}). It means that the procedure to initiate the movement should be as follows: for $\alpha < 3/4$ the base of the cone should be raised near the free surface of the liquid; for $\alpha > 3/4$ the vertex of the cone should be lowered until near complete immersion; for $\alpha = 3/4$ the choice of the base or vertex to initiate the movement is irrelevant since the amplitude of the oscillation, that has a maximum (equal to $h$) corresponds to a displacement between the base and the vertex i.e.\ $\overline{y}\in [-\overline{y}_{\mbox{\scriptsize C}}, -\overline{y}_{\mbox{\scriptsize V}}]$. This can be seen in Fig.\,(\ref{fig:FigureDoisSemCor}).

\begin{figure}[h!]
\centering
\includegraphics[width=4.0 in]{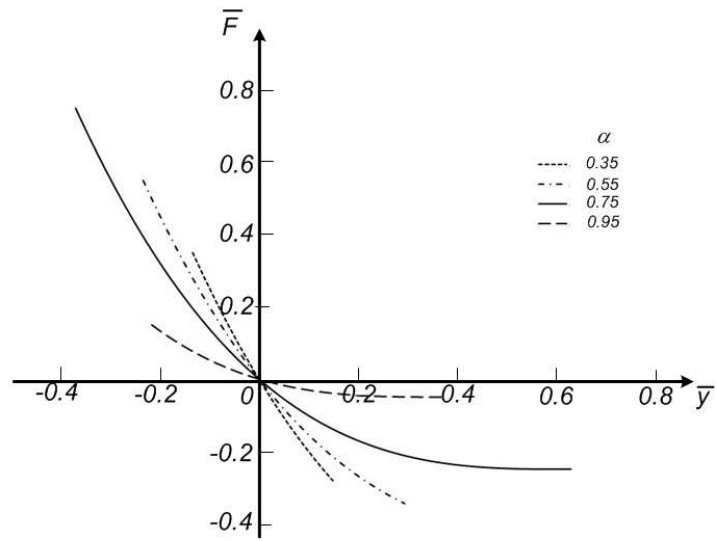}
\caption{The reduced restoring force as a function of the reduced displacement, $\overline{F}=f(\overline{y})$ , within the proper interval of oscillation taking $\alpha$ as a parameter.}
\label{fig:FigureTresSemcor}
\end{figure}

\begin{figure}[h!]
\centering
\includegraphics[width=4.0 in]{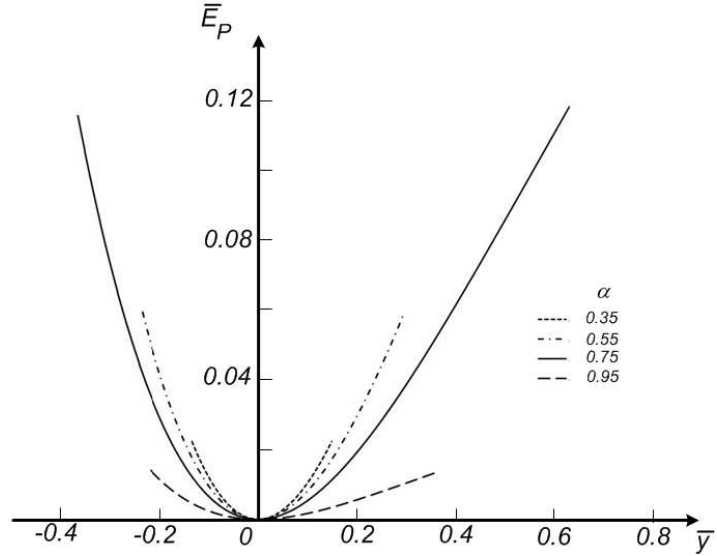}
\caption{The reduced potential energy $\overline{E}_{\mbox{\scriptsize P}}=f(\overline{y})$, within the proper interval of oscillation taking $\alpha$ as a parameter.}
\label{fig:FigureQuatroSemcor}
\end{figure}

The restoring force (reduced value) $\overline{F}=F/F_0=-P_3(\overline{y})$ and the potential energy $\overline{E}_{\mbox{\scriptsize P}}=E_{\mbox{\scriptsize P}}/E_0=P_4(\overline{y})$ can be represented in the interval of oscillation with $\alpha$ as parameter [Fig.\,(\ref{fig:FigureTresSemcor}) and Fig.\,(\ref{fig:FigureQuatroSemcor})].

\section{General discussion of the type of oscillation}

\begin{figure}[h!]
\centering
\includegraphics[width=4.2 in]{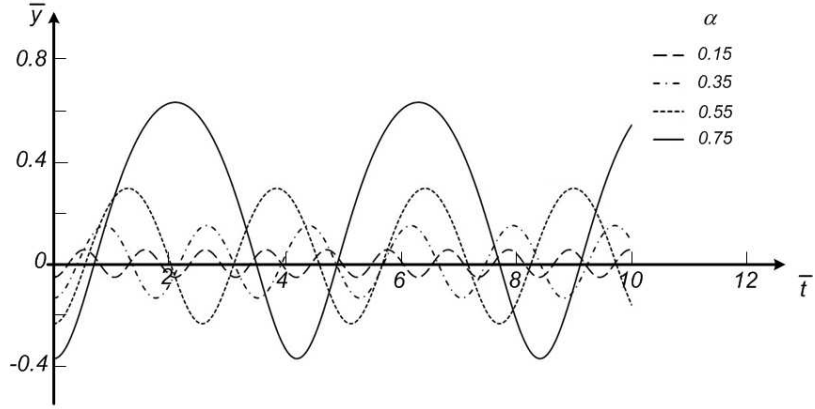}
\caption{Displacement of the cone (reduced values) as a function of the reduced time for several values of the parameter $\alpha\leq 3/4$.}
\label{fig:figuraCinco}
\end{figure}

\begin{figure}[h!]
\centering
\includegraphics[width=4.0 in]{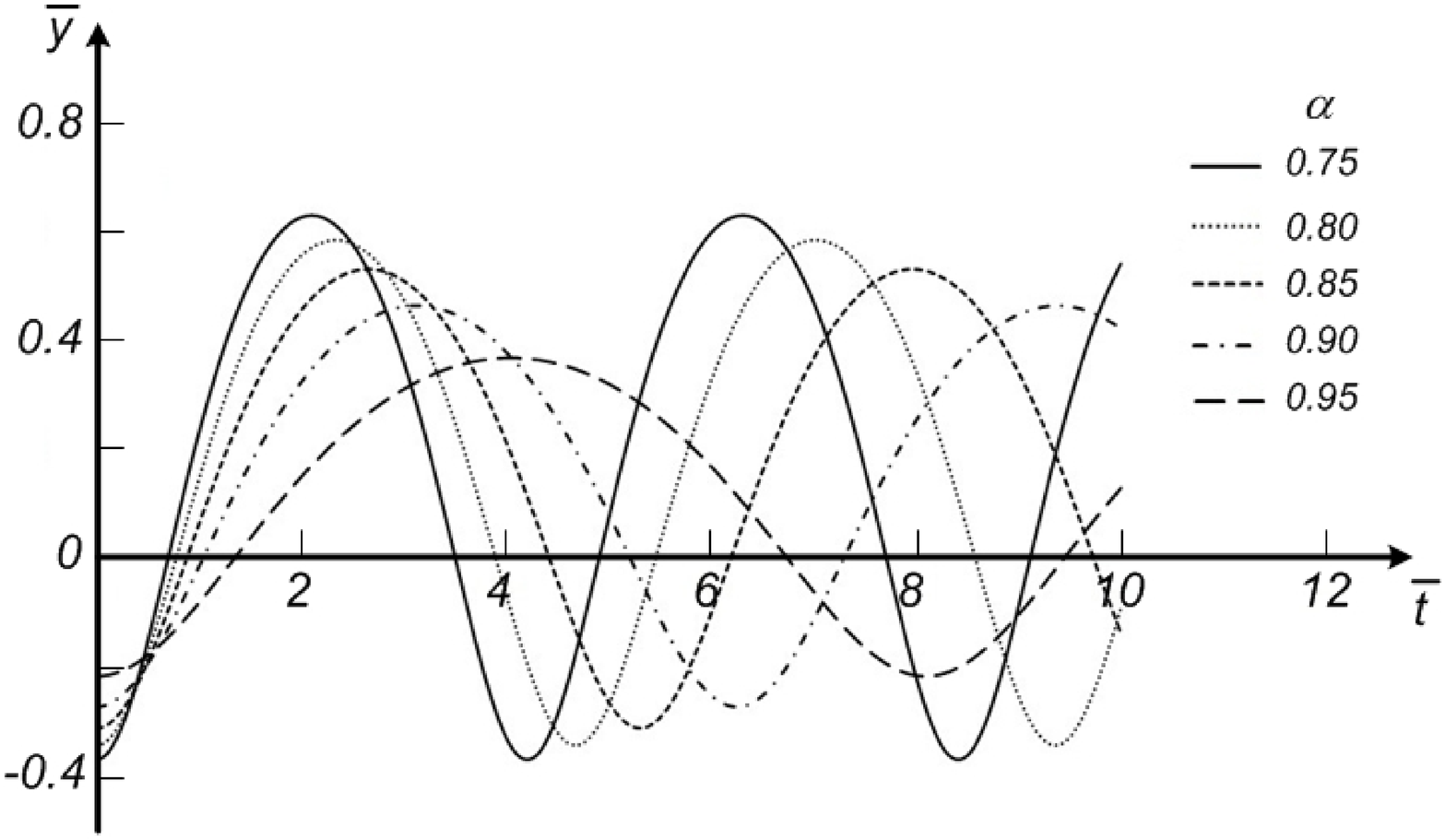}
\caption{Displacement of the cone (reduced values) as a function of the reduced time for several values of the parameter $\alpha\geq 3/4$.}
\label{fig:figuraSeis}
\end{figure}

\begin{figure}[h!]
\centering
\includegraphics[width=4.0 in]{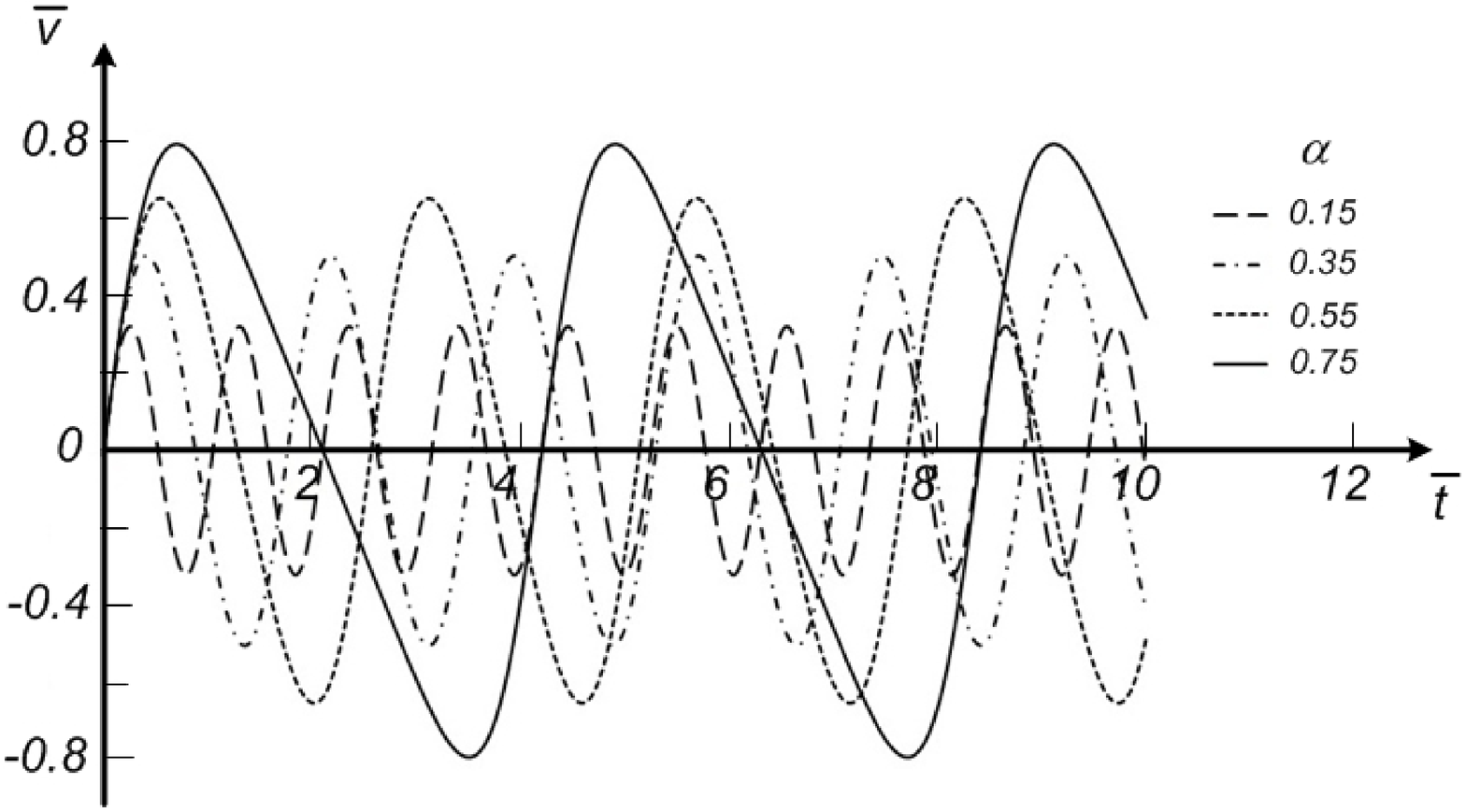}
\caption{Reduced value of the velocity as a function of the reduced time for several values of the parameter $\alpha\leq 3/4$.}
\label{fig:figuraSete}
\end{figure}

\begin{figure}[h!]
\centering
\includegraphics[width=4.0 in]{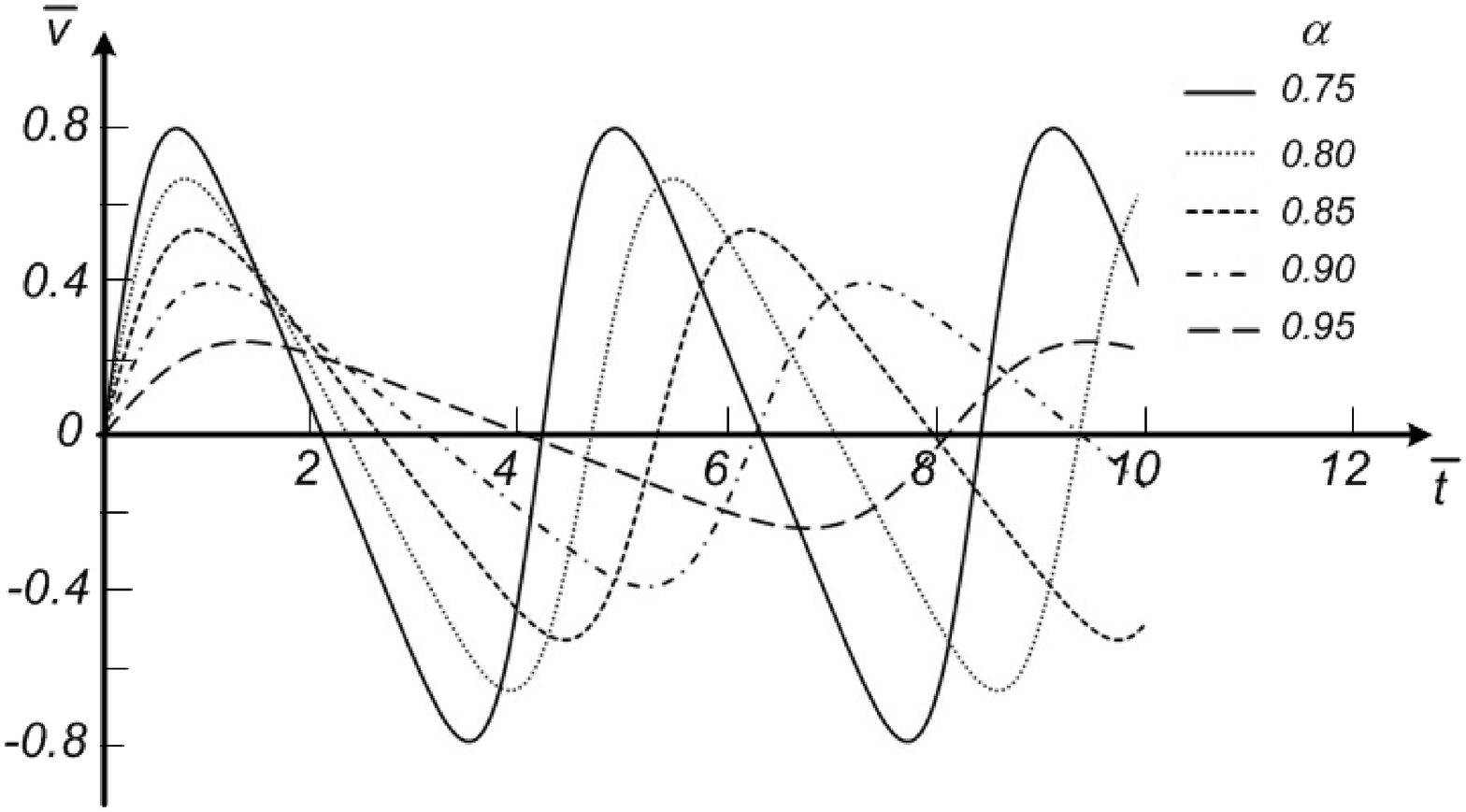}
\caption{Reduced value of the velocity as a function of the reduced time for several values of the parameter $\alpha\geq 3/4$.}
\label{fig:figuraOito}
\end{figure}

We use the Newton equation to find the position of the point $P$ of the cone as a function of time. The adoption of reduced values continues to offer simplification in the final equation. Let us define the unit of time $t_0=\sqrt{2\,(h/g)}$. Therefore $\overline{t}=t/t_0\Longrightarrow dy/dt=(h/t_0)\,d\overline{y}/d\overline{t}$. The unit for $v$ is $v_0=\sqrt{gh/2}$ which points to $\overline{v}=v/v_0=d\overline{y}/d\overline{t}$. On the other hand $d^2y/dt^2=(h/t_0^2)\,d^2\overline{y}/d\overline{t}^2$, i.e.\ the equation of motion in terms of the reduced variables is 
\begin{equation*}
\rho V_{\mbox{\scriptsize C}} \frac{h}{t_0^2}\,\frac{d^2\overline{y}}{d\overline{t}^2}=F \Longleftrightarrow \frac{d^2\overline{y}}{d\overline{t}^2}= \frac{t_0^2}{h}\,\frac{F}{\rho_{\mbox{\scriptsize f}}\,\alpha V_{\mbox{\scriptsize C}}} \Longleftrightarrow \frac{d^2\overline{y}}{d\overline{t}^2}=\frac{2}{\alpha}\,\frac{F}{F_0}.
\end{equation*}
Finally 
\begin{equation*}
\frac{d^2\overline{y}}{d\overline{t}^2}-\frac{2}{\alpha}\,\overline{F}=0 \Longleftrightarrow \frac{d^2\overline{y}}{d\overline{t}^2}+\frac{2}{\alpha} P_3=0.
\end{equation*}
The system does not oscillate symmetrically about the origin since the force (polynomial $P_3$) shows odd and even powers. For small displacements $\overline{y}\ll 1,\, P_3(\overline{y})\approx 3 (1-\alpha)^{2/3}\overline{y}$, the equation is approximately linear and so the movement is quasi harmonic with period $\overline{T}_{SO}=2\pi \sqrt{\alpha/6}\,\bm .\, \sqrt[3]{1/(1-\alpha)}$. In other physical situations the oscillator exhibits anharmonic displacements and the period will depend on the amplitude (or total energy).
The solutions $\overline{y}=\overline{y}(\overline{t})$ and $\overline{v}=\overline{v}(\overline{t})$ were obtained by computational methods and are represented in Figs. (\ref{fig:figuraCinco}), (\ref{fig:figuraSeis}), (\ref{fig:figuraSete}) and (\ref{fig:figuraOito}).

\section{The Period}

The conservation of the energy of the oscillator allows the determination of  the velocity as a function of displacement: $d\overline{v}/d\overline{t}+(2/\alpha) P_3=0 \Longrightarrow \overline{v}\,d\overline{v}=-(2/\alpha) P_3\,d\overline{y} \Longrightarrow \overline{v}=\pm (2/\sqrt{\alpha})\,\sqrt{\overline{E}_{total}-P_4}$.

\begin{equation*}\small
\hspace{-24mm}\overline{v}=\pm\left\{
\begin{array}{ll}
\frac{1}{\sqrt{\alpha}}\,\sqrt{3(1-\alpha)^{4/3}+4\alpha-3-\overline{y}^4+4(1-\alpha)^{1/3}\overline{y}^3-6(1-\alpha)^{2/3}\overline{y}^2}&\Longleftarrow \alpha<\frac{3}{4}\\\vspace{-3mm}
\\
\frac{1}{\sqrt{\alpha}}\,\sqrt{3(1-\alpha)^{4/3}-\overline{y}^4+4(1-\alpha)^{1/3}\overline{y}^3-6(1-\alpha)^{2/3}\overline{y}^2}&\Longleftarrow \alpha\geq\frac{3}{4}.
\end{array}\right.
\end{equation*}

\begin{figure}[h!]
\centering
\includegraphics[width=4.0 in]{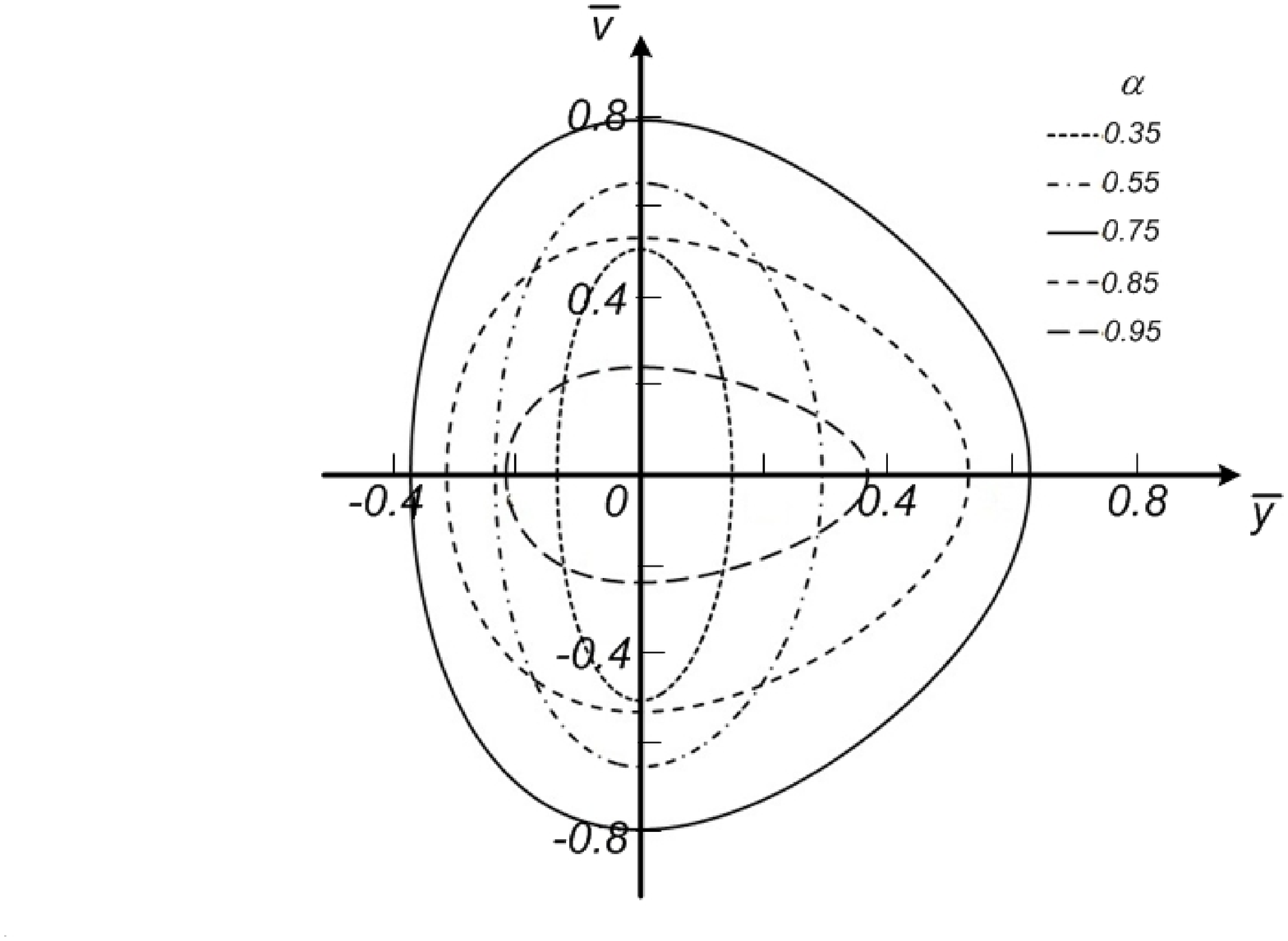}
\caption{The phase plane representation ($\overline{v}$ as a function of $\overline{y}$) for some values of $\alpha$.}
\label{fig:NovaFiguraNove}
\end{figure}

The phase space representation is shown in Fig. (\ref{fig:NovaFiguraNove}). From the symmetry of these curves relatively to the horizontal axis, we can conclude that the interval of time between two zeros of the velocity or two extremes of the displacements is equivalent to half of the period of the movement. Since $d\overline{t}=d\overline{y}/\overline{v}$, $\overline{T}/2=\int_a^b d\overline{t}$. Then,
\begin{equation*}
\frac{\overline{T}}{2}=\left\{
\begin{array}{ll}
\displaystyle\int\limits_{-\overline{y}_{\mbox{\scriptsize C}}}^{\overline{y}_1}\frac{d\overline{y}}{\overline{v}}&\Longleftarrow \alpha<\frac{3}{4}\\\vspace{-3mm}
\\
\displaystyle\int\limits_{\overline{y}_1}^{-\overline{y}_{\mbox{\scriptsize V}}}\frac{d\overline{y}}{\overline{v}}&\Longleftarrow \alpha\geq\frac{3}{4}.
\end{array}\right.
\end{equation*}

\begin{figure}[h!]
\centering
\includegraphics[width=4.0 in]{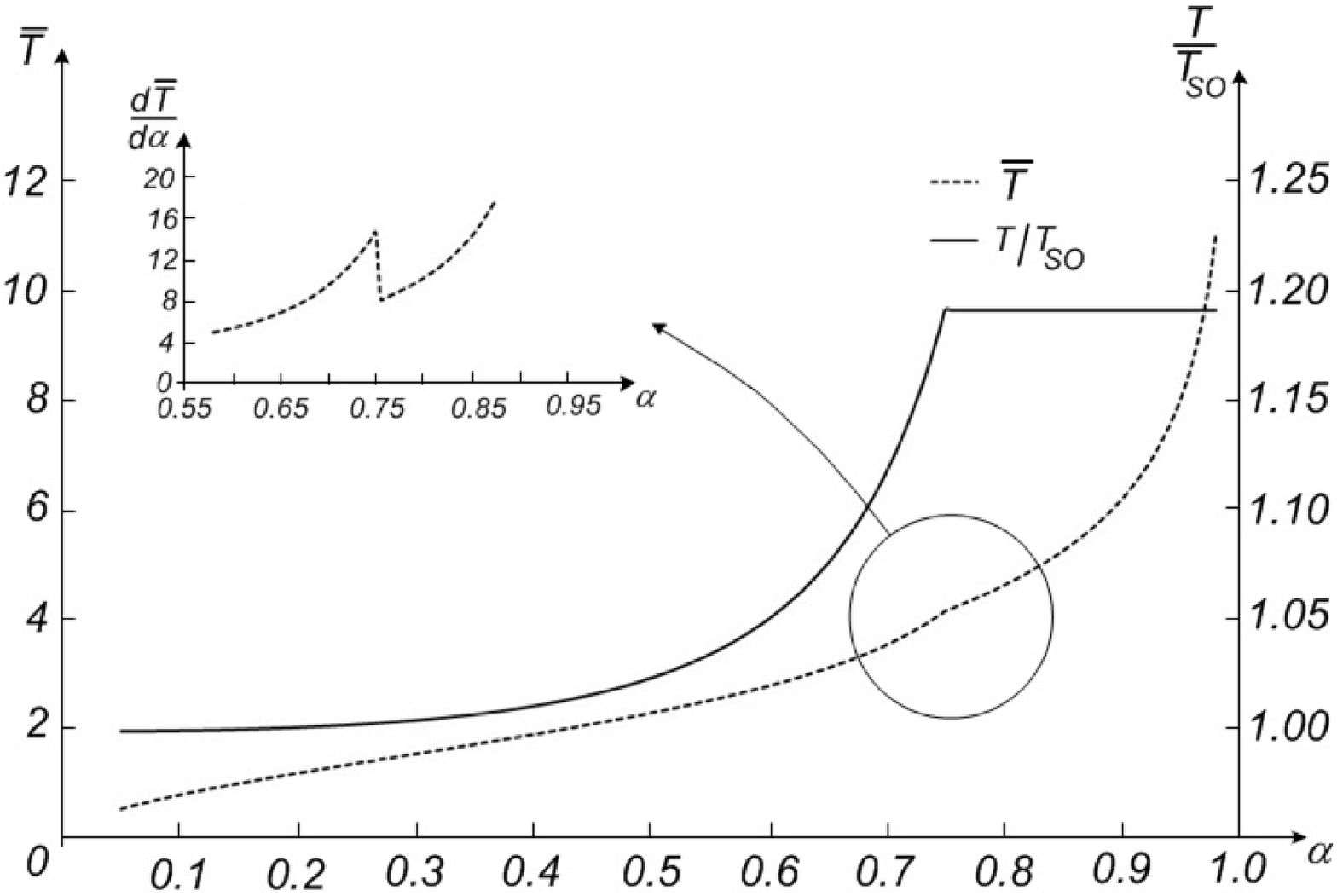}
\caption{The reduced (left axis) and normalized (right axis) period as a function of $\alpha$, calculated for the highest value of its total energy. The derivative $d\overline{T}/d\alpha$ is also shown in the inset.}
\label{fig:FiguraDezPretoBranco}
\end{figure}

The limits of integration and the integrand function have been determined above. The result is only a function of $\alpha$.
It is possible to get a close solution for the period in terms of complete elliptic integrals of the first kind\cite{lebev}. For $\alpha<3/4$, the analytical solution is far too complicated to be considered and so numerical methods were used. However, the solution for $\alpha\geq 3/4$ was easily obtained: 
\begin{eqnarray*}\nonumber
\overline{T}=&2\sqrt{\alpha}\int\limits_{-(\sqrt[3]{4}-1)\sqrt[3]{1-\alpha}}^{\sqrt[3]{1-\alpha}}\,\frac{d\overline{y}}{\sqrt{3(1-\alpha)^{4/3}-\overline{y}^4+4(1-\alpha)^{1/3}\overline{y}^3-6(1-\alpha)^{2/3}\overline{y}^2}}\\\vspace{-4mm}
\\\nonumber
\overline{T}=&\frac{\sqrt{\alpha}}{\sqrt[3]{1-\alpha}}\,\frac{2^{4/3}}{3^{1/4}}~K\hspace{-1mm}\left(\frac{\sqrt{2}}{2(1+\sqrt{3})}\right)=C\frac{\sqrt{\alpha}}{\sqrt[3]{1-\alpha}},
\end{eqnarray*}
where $K\left(\sqrt{2}/(2(1+\sqrt{3}))\right)$ is a complete elliptic integral of the first kind of argument $\sqrt{2}/(2(1+\sqrt{3}))$.
The numerical value of  C is: $C=3.059908075$. An interesting result that should be outlined is the fact that the normalized value of the period for $\alpha\geq 3/4$, 
$T_{norm}=\overline{T}/\overline{T}_{SO}$ ($\overline{T}_{SO}$, period of small oscillations) is a constant, i.e.\ does not depend on $\alpha$:
\begin{equation*}
T_{norm}=C\frac{\sqrt{\alpha}}{\sqrt[3]{1-\alpha}}\,\frac{\sqrt{6}\sqrt[3]{1-\alpha}}{2\pi\,\sqrt{\alpha}}=\frac{\sqrt{6}}{2\pi}C= 1.192900269.
\end{equation*} 
The period of the cone oscillations at the highest energy as a function of $\alpha$ is represented in Fig. (\ref{fig:FiguraDezPretoBranco}) together with the normalized values. 

\begin{figure}[h!]
\centering
\includegraphics[width=4.0 in]{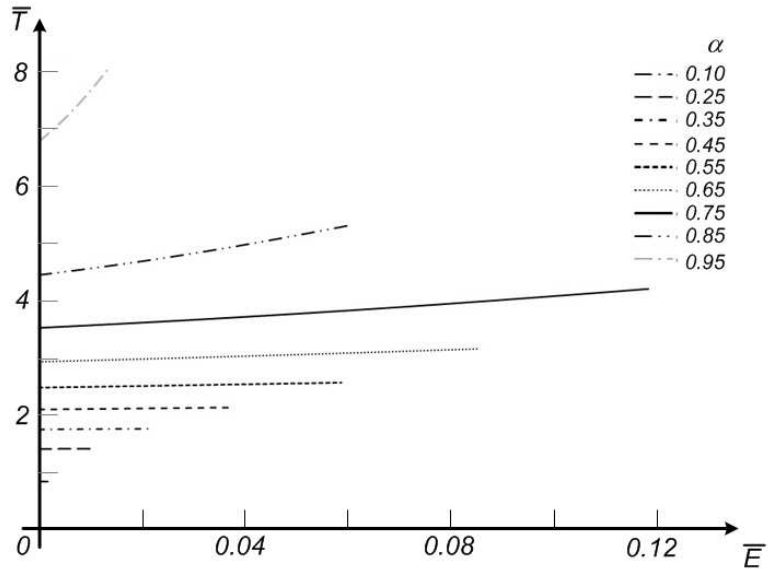}
\caption{The variation of the period with the energy (reduced values) using $\alpha$ as a parameter. The energy varies from $0$ to its maximum value {\large(}Eq.\,(\ref{eqyzero}){\large)}.}
\label{fig:FigureOnzeSemcor}
\end{figure}
So far, we have represented the oscillation with the highest energy or maximum value of the initial amplitude satisfying conditions (\ref{conditons}). It is interesting to examine closely the dependence of the period with the energy of the oscillator from zero to the highest value given by Eqs.\,(\ref{eqyzero}). This is shown in Fig.\,(\ref{fig:FigureOnzeSemcor}) taking, as usual $\alpha$ as a parameter. It is interesting to observe that the numerically obtained values ($\overline{E}, \overline{T}$) fit $100\,\%$ a second degree polynomial.

\begin{figure}[h!]
\centering
\includegraphics[width=3.5 in]{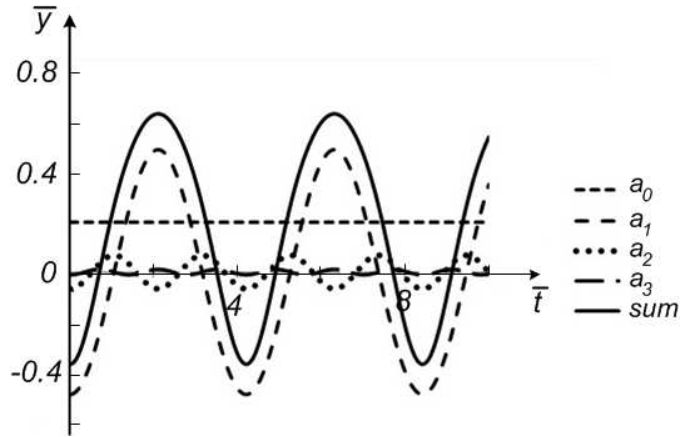}
\caption{Harmonic components of the reduced displacement for $\alpha=3/4$.}
\label{fig:FigureDozeSemcor}
\end{figure}

\section{Fourier analysis.}

The Fourier analysis is a very useful tool to investigate the harmonic components of a periodic function of time that derives from a non linear equation, as it happens in our study. From Figs.\,(\ref{fig:figuraCinco}) and (\ref{fig:figuraSeis}) one can conclude that the displacement $\overline{y}=\overline{y}(\overline{t})$ is a periodic symmetric even function with a non null mean value. The Fourier decomposition gives for the general case: $y(t)=a_0+\sum\limits_{n=1}^{\infty}a_n\cos{\left(n\frac{2\pi}{T}\,\right)}+\sum\limits_{n=1}^{\infty}b_n\sin{\left(n\frac{2\pi}{T}\,\right)}$; in this case all $b_n=0$ since $y(t)$ is an even function. The coefficients $a_n$ were calculated by numerical methods once the fundamental frequency is known. Only the first four coefficients were significant. 
The Fourier components for the particular case $\alpha=\frac{3}{4}$ that corresponds to the maximum of the highest value of energy are represented in Fig.\,(\ref{fig:FigureDozeSemcor}). It also exhibits the highest coefficients which indicate the highest degree of anharmonicity. 
The reduced values of all calculated coefficients ($n=0,...,3$) are also shown in Table\,\ref{table}.
\begin{table}[h]\footnotesize
\caption[]{Values of the coeficients $a_n$ ($n=0,\dots, 3$) of the Fourier analysis of the reduced displacement, $\overline{y}=\overline{y}(\overline{t})$.}\label{table}
\begin{tabular}{llllllll}
\br
{$\bm \alpha$}&{$\bm E_{total}$}&{$\bm y_0$}&{$T$}&{$\bm a_0$}&{$\bm a_1$}&{$\bm a_2$}&{$\bm a_3$}\\
\mr
$0.15$&$0.003884$&$-0.052732$&$1.04975$&$1.525\,10^{-3}$&$-5.374\,10^{-2}$&$-5.08\,10^{-4}$&$-5.39\,10^{-6}$\\
$0.25$&$0.011065$&$-0.091440$&$1.41611$&$4.941\,10^{-3}$&$-9.470\,10^{-2}$&$-1.65\,10^{-3}$&$-3.23\,10^{-5}$\\
$0.35$&$0.022292$&$-0.133761$&$1.76556$&$1.157\,10^{-2}$&$-1.413\,10^{-1}$&$-3.86\,10^{-3}$&$-1.19\,10^{-4}$\\
$0.45$&$0.037970$&$-0.180679$&$2.13573$&$2.359\,10^{-2}$&$-1.960\,10^{-1}$&$-7.89\,10^{-3}$&$-3.60\,10^{-4}$\\
$0.55$&$0.058629$&$-0.233691$&$2.57014$&$4.570\,10^{-2}$&$-2.630\,10^{-1}$&$-1.53\,10^{-2}$&$-1.02\,10^{-3}$\\
$0.65$&$0.084992$&$-0.295270$&$3.15971$&$8.998\,10^{-2}$&$-3.515\,10^{-1}$&$-3.03\,10^{-2}$&$-3.02\,10^{-3}$\\
\mr
$\bm{0.75}$&$ 0.118118$&$ -0.370039$&$ 4.20655$&$ 1.988\,10^{-1}$&$ -4.884\,10^{-1}$&$ -6.69\,10^{-2}$&$ -1.13\,10^{-2}$\\
\mr
$0.85$&$0.059775$&$-0.312103$&$5.30951$&$1.677\,10^{-1}$&$-4.120\,10^{-1}$&$-5.65\,10^{-2}$&$-9.50\,10^{-3}$\\
$0.95$&$0.013815$&$-0.216400$&$8.0956$&$1.162\,10^{-1}$&$-2.856\,10^{-1}$&$-3.92\,10^{-2}$&$-6.58\,10^{-3}$\\
\br
\end{tabular}
\end{table}

\section{Conclusions}
Most real oscillators contain anharmonic components. The present study of a floating cone movement is by all means an interesting case of an anharmonic oscillator. The fact that the restoring force is a polynomial of the third degree with non null coefficients (except the independent term) imply an asymmetry of this force and the related potential curve.
Comparing this  to well known based mass-spring systems results that the equivalent spring stiffness is no longer constant. It varies as a sum of two contributions that have opposite sign from a specific position. In solid state physics we encounter these kind of forces such as the cohesive force containing a short range repulsion (hard sphere interaction) and a long range attraction. The former varies with the displacement much faster than the latter.

\end{document}